\documentclass[twocolumn,a4paper]{article}  
\usepackage{graphicx}
\usepackage{amssymb}
\usepackage{amsmath}
\usepackage{epstopdf}
\usepackage{array}
\newcolumntype{L}[1]{>{\raggedright\let\newline\\\arraybackslash\hspace{0pt}}m{#1}}
\newcolumntype{C}[1]{>{\centering\let\newline\\\arraybackslash\hspace{0pt}}m{#1}}
\newcolumntype{R}[1]{>{\raggedleft\let\newline\\\arraybackslash\hspace{0pt}}m{#1}}
\usepackage[table]{xcolor}
\usepackage{booktabs}
\usepackage{multirow}
\usepackage[colorlinks,citecolor=red,urlcolor=blue,bookmarks=false,hypertexnames=true]{hyperref} 
\usepackage{doi} 
\colorlet{tableheadcolor}{gray!25}
\colorlet{tablerowcolor}{gray!12.5}
\usepackage{tablefootnote}
\usepackage{fvextra}
\usepackage[frozencache=true,cachedir="./"]{minted}
\usepackage{siunitx}
\usepackage{lipsum}

\title{Database support of detector operation and data analysis in the DEAP-3600 Dark Matter experiment.}

\begin{document}
\author{T.R. Pollmann$^1$, B. Smith$^2$  }    

\date{%
$^1$ Department of Physics E15, Technische Universit\"at M\"unchen, James-Franck-Str. 1, 85748 Garching, Germany \\
$^2$ TRIUMF, Vancouver V6T 2A3, Canada\label{addr2}
}

\maketitle

\begin{abstract}
The DEAP-3600 detector searches for dark matter interactions on a \SI{3.3}{tonne} liquid argon target. Over nearly a decade, from start of detector construction through the end of the data analysis phase, well over 200 scientists will have contributed to the project. The DEAP-3600 detector will amass in excess of \SI{900}{TB} of data representing more than 10$^{10}$ particle interactions, a few of which could be from dark matter. At the same time, metadata exceeding \SI{80}{GB} will be generated. This metadata is crucial for organizing and interpreting the dark matter search data and contains both structured and unstructured information.

The scale of the data collected, the important role of metadata in interpreting it, the number of people involved, and the long lifetime of the project necessitate an industrialized approach to metadata management.

We describe how the Couch\-DB and the PostgreSQL database systems were integrated into the DEAP detector operation and analysis workflows. This integration provides unified, distributed access to both structured (PostgreSQL) and unstructured (Couch\-DB) metadata at runtime of the data analysis software. It also supports operational and reporting requirements.




\end{abstract}

\tableofcontents

\section{Introduction}
Rare event searches today are undertaken by collaborations of $\mathcal O (100)$ researchers who construct large detectors and operate them for many years. The scale and the duration of the projects makes database approaches necessary for managing information about the detector components and the datasets. Projects have, for example, taken a database approach to managing the quality of detector components and their status~\cite{Loach:2016bo,Collaboration:2015if,Golling:2012ct}, and to keeping track of datasets~\cite{Cox:2012zz,Liu:2014hg} and detector conditions~\cite{Gruttola:2010dma}. In this work, we present a unified database approach to keeping track of detector hardware, response parameters, environment information, and datasets for the DEAP-3600 Dark Matter detector~\cite{detectorpaper,Amaudruz:2018gr,Ajaj:2019wi}.

Dark matter particles would create a very specific light signature when they scatter on argon nuclei in the DEAP-3600 detector~\cite{Boulay:2009ug}. This signature is searched for in the waveform data from 255~photomultiplier tubes (PMTs), which observe scintillation light from \SI{3.3}{tonnes} of liquid argon (LAr) for nominally 4~years. Calibrations of the empty detector started in 2015. The detector was filled with LAr and started taking dark matter search data (so-called \emph{physics data}) at the end of 2016.

Most of the recorded waveforms are not from dark matter interactions, because the dark matter interaction probability is extremely small. In fact, the signals from such an interaction will be hidden beneath the signals from over 10$^{10}$ scintillation events caused by the decay of natural radioactive isotopes in the detector. In order to interpret the PMT waveforms correctly and thus to reliably identify the background events, the detector status and response properties over the 5.5~year operation period must be recorded, and made accessible to analysis. This information is typically referred to as \emph{non-event data} in particle physics, or generally as \emph{metadata}.

We implemented a central store for non-event data based on remote database servers running the Couch\-DB and PostgreSQL database systems. Access to non-event data is mediated through HTTP requests from either the project's C++/Python analysis framework or the project website. Through the databases, we ensure that crucial information about the detector and the data is permanently and easily available to analyzers, that everyone uses the same, most up-to-date analysis inputs, and that these inputs do not change without a record.

In Section~\ref{sect:datataking} we describe briefly how physics data is collected and analysed, as this informs the requirements on the non-event data stores. For a detailed description of the DEAP detector and the triggering scheme we refer the reader to Ref.~\cite{detectorpaper}. The physical setup of the database servers is explained in Section~\ref{sect:infrastructure}. Section~\ref{sect:CouchDB} describes the Couch\-DB and Section~\ref{sect:dpostgres} the PostgreSQL implementation. Some of the workflows enabled by the database systems are described in Section~\ref{sect:workflow}. Finally, the performance of the database systems is presented in Section~\ref{sect:performance}.

\section{Data taking and analysis in DEAP} \label{sect:datataking}

\subsection{Data streams} \label{subsect:datastreams}

The DEAP experiment has three data streams that differ by orders of magnitude in acquisition rate, shown in Tab.~\ref{tab:daqchannels}.

\begin{table}[htp]
\caption{Overview of the data acquisition streams, data volumes, and formats, used in DEAP-3600. The last two columns are the topic of this work. In the \emph{purpose} row, ``science'' means dark matter search data, ``veto'' means that information therein is used to discard or \emph{veto} some part of the the science data, and ``diagnostics'' means the data therein is used to check that the detector is working properly. }
\begin{center}
\rowcolors{3}{tablerowcolor}{}
\begin{tabular}{L{0.14\columnwidth}|L{0.22\columnwidth}|L{0.2\columnwidth}|L{0.2\columnwidth}}
\rowcolor{tableheadcolor}
           &fast DAQ (event-based) &slow control (time-based) & metadata (run-based) \\
  Purpose   & science, veto &veto, diagnostics &veto, calibration, diagnostics  \\
  Channels & 562 & 300 & NA  \\
  Rate &  50--250$\cdot 10^6$ S/s/channel & $\sim$1 S/s/channel & $\sim$50 documents/day \\
  Data volume & $\sim$600 GB/day & $\sim$30 MB/day & $\sim$5 MB/day \\
  Format &ROOT (B-Tree) & PostgreSQL & Couch\-DB \\
\end{tabular}
\end{center}
\label{tab:daqchannels}
\end{table}%

The \emph{fast DAQ} stream consists of waveforms read out from 255 LAr-facing PMTs and 52 veto PMTs in response to scintillation events occuring at a rate of approximately \SI{3300}{\hertz}. When a trigger condition is met, \SI{16}{\micro\second} long waveforms are digitized by the data acquisition system (DAQ) at 250 million samples per second (MS/s) and at \SI{62.5}{MS/s}  (2 read-out channels for each LAr-facing PMT) and saved for offline analysis in the ROOT~\cite{root} particle physics data format. Such a set of up to $255\cdot2 + 52$ digitized waveforms is called an \emph{event}. An assembly of events recorded in the same DAQ configuration for a continuous stretch of time is called a \emph{run}. The data is filtered and compressed online, and not all channels are digitized on each trigger, hence the amount of data written to disk is much smaller than a naive calculation would indicate.

The other two data-streams comprise the non-event data this work is concerned with.

The \emph{slow control} system records environmental data such as the temperatures, pressures, and liquid levels in detector sub-systems every 1--2~seconds. This data is stored internally on the commercial DeltaV control system~\cite{deltaV}. The data from 127 sensors that affect the interpretation of physics data are continuously exported to a PostgreSQL database. 

Any other metadata that is needed in any way to support physics data analysis is stored in Couch\-DB. The objects whose properties are tracked, sorted by category, are listed in Tab.~\ref{tab:tagnames}.

\begin{table}[!htbp]
\caption{Objects whose properties and relationships are tracked in Couch\-DB. Objects are distinguished by alpha-numeric tags. We give names to tags belonging to the same category to simplify referencing these later. The grouping of objects into three databases within Couch\-DB will be explained in Section~\ref{sect:infrastructure}.}
\begin{center}
\rowcolors{3}{tablerowcolor}{}
\begin{tabular}{L{0.32\columnwidth}|L{0.22\columnwidth}|L{0.18\columnwidth}|L{0.12\columnwidth}}
\rowcolor{tableheadcolor}
 Category   & Tag name & Number of objects \tablefootnote{Where numbers are approximate, they are the sum of the 1.5 year calibration time and an extrapolation from the current 2.5 to a total 4 year physics run-time.} & DB \\
 \hline
 PMT   & PMTID & 307 &\cellcolor{white!0}  \\ 
 Calibration source & SourceID & 29 &\cellcolor{white!0} \\
 Slow control sensor & sensor tag & 150 &\cellcolor{white!0}  \\ 
 Run   & RunID & $\sim$20000 &\cellcolor{white!0} \\
 Group of runs that belong to the same analysis & Runlist & $\sim$340 &\cellcolor{white!0} \\ 
 Data quality question & QuestionID & $\sim$50 &\cellcolor{white!0} \\
 Channel on DAQ boards (digitizers, signal conditioners, HV supply) & ChannelID & 865 & \cellcolor{white!0}\multirow{-7}{*}{deap} \\
 \hline
 DAQ operator & FirstnameLastname & $\sim$200 & \cellcolor{white!0}\\
 DAQ shift & shift start date & $\sim$1725 &  \cellcolor{white!0}\multirow{-2}{*}{schedule}  \\ 
 \hline
 DAQ settings   & RunType & $\sim$110 & daq \\ 
\end{tabular}
\end{center}
\label{tab:tagnames}
\end{table}%





\begin{figure}[htbp]
\begin{center}
\includegraphics[width=1.0\columnwidth]{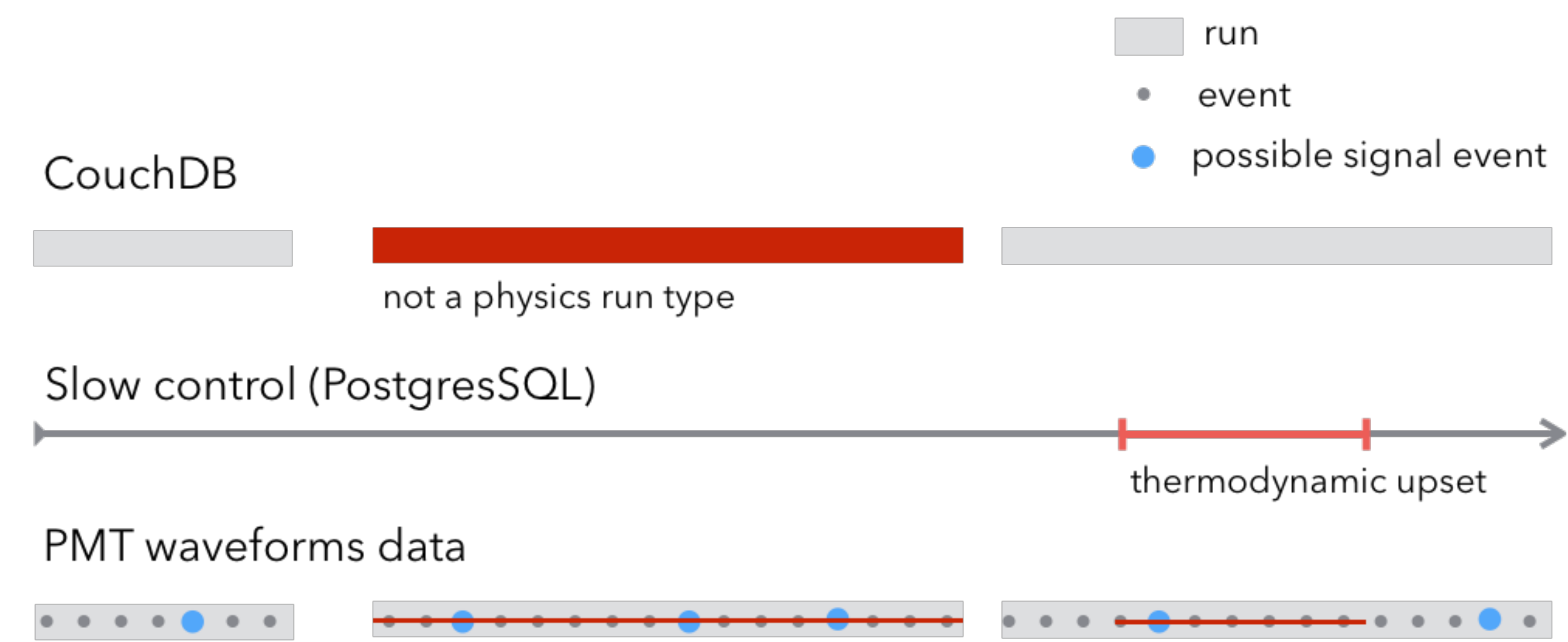}
\caption{Schematic diagram of the data selection workflow. The Couch\-DB database is checked for information on the runs. Certain runs are discarded for example because they are not a physics run type (they may be a calibration run) or because they were flagged as problem runs (for example because a DAQ subsystem crashed). The thermodynamic conditions of the detector are then checked and events during upset conditions are discarded. Finally, possible dark matter signal events are identified out of all remaining recorded events.}
\label{fig:vetoconditions}
\end{center}
\end{figure}

The interaction and relationship between the data streams is illustrated in Fig.~\ref{fig:vetoconditions} using data selection as an example. 

\subsection{The RAT analysis tool}

Physics data is saved to disk and reduced offline using a C++/Python analysis framework custom written for the analysis of scintillation signals, called RAT~\cite{RAT}. The RAT architecture is used by several collaborations, such as MiniCLEAN~\cite{Hime:2011tt} and SNO+~\cite{Andringa:2015tza}. The RAT-DEAP branch of the codebase was specifically developed and extended for the DEAP experiment.

The default RAT architecture contains functionality to read local JSON-like documents at run-time, and defines a structure for these JSON documents so that they can be used for database-like lookups. The JSON format is an unordered collection of key/value pairs, stored as human-readable text. Local JSON files are used to define material properties and detector geometry for simulation purposes, and store settings that determine how the code behaves. 

In RAT-DEAP we have kept the conceptual database (DB) architecture envisioned in the original RAT codebase. We describe here an implementation of the Couch\-DB and PostgreSQL backends as well as a number of extensions to RAT which enable revision control and the storing of information by RunID.

\section{Database infrastructure} \label{sect:infrastructure}

\begin{figure}[htbp]
\begin{center}
\includegraphics[width=\columnwidth]{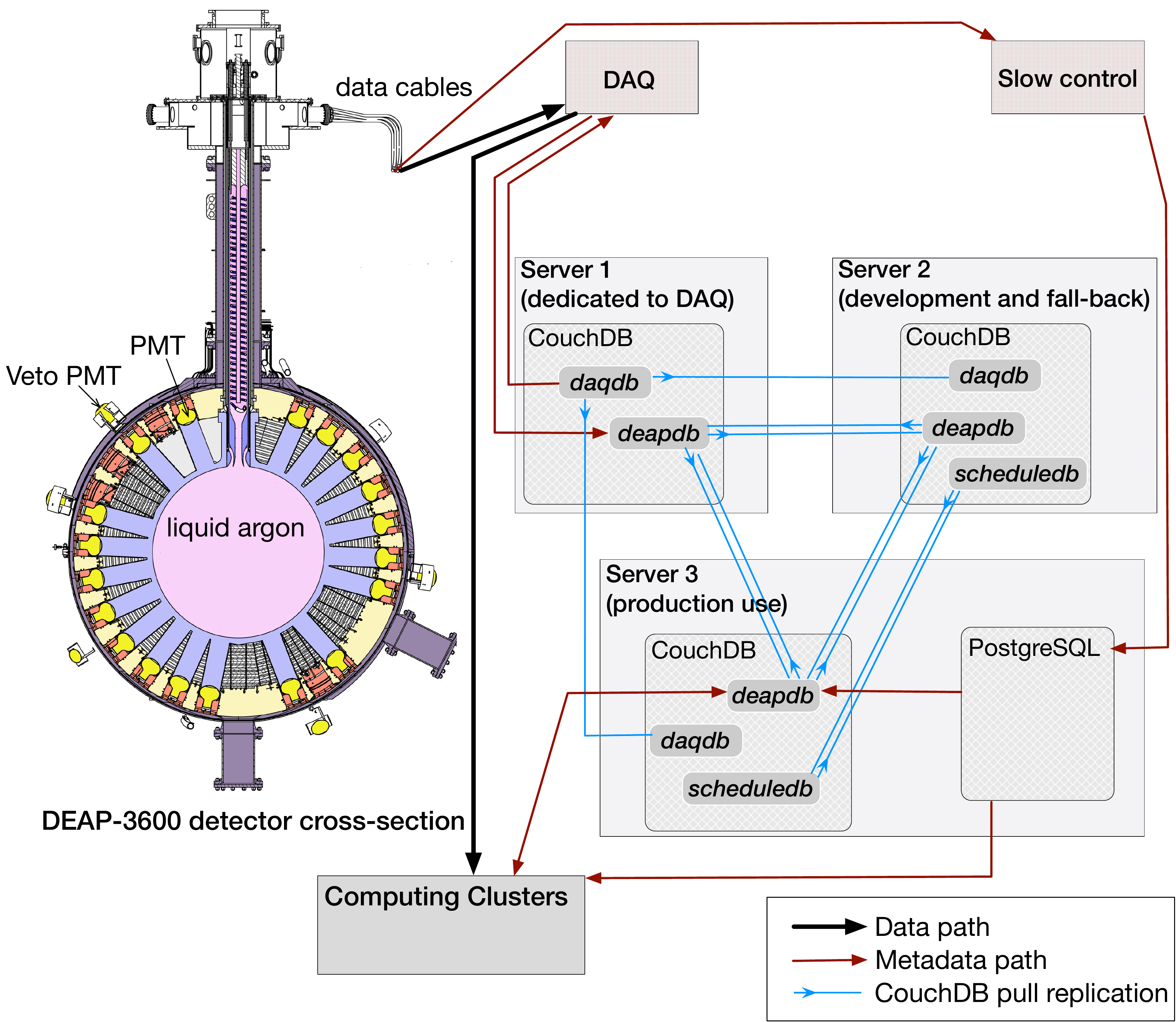}
\caption{The flow of event and non-event data between the DEAP-3600 detector and different computing systems.}
\label{fig:dbrouting}
\end{center}
\end{figure}

Fig.~\ref{fig:dbrouting} is a schematic representation of the detector hardware and computing systems. The detector is located at SNOLAB, in  a mine \SI{2}{km} underground. The DAQ and slow control systems are physically next to the detector. 

Three servers run an instance of the DEAP Couch\-DB. Server~1 is dedicated to DAQ operation and is near the DAQ racks. This allows the DAQ system to take data and write metadata even when the network connection is interrupted. Server~2, located on surface, is used for development of new interfaces or data structures and for resource-intensive queries. Server~3 is used for regular user queries. It is located off-site to protect against site-specific downtime.

Within Couch\-DB, information is stored in three separate databases. The objects tracked in each database are listed in Tab.~\ref{tab:tagnames}. The main physics-analysis database is \emph{deapdb}. The RAT-DEAP analysis framework relies on this database for detector status and calibration information. The other two databases support operational tasks. 
 \emph{daqdb} is solely used by the DAQ system to manage the DAQ settings. The DAQ accesses this database by direct HTTP requests. \emph{scheduledb} serves as the database backend to a webpage through which DAQ shift sign-up and scheduling are managed. It is usually accessed from this webpage. Separating information into different databases within the Couch\-DB system simplifies access control and database management tasks. For example, maintenance or improvements can be done on \emph{scheduledb} without impacting access to information stored in \emph{deapdb}.

Each server runs standard Couch\-DB continuous pull replications to synchronise its databases. 

All user queries by default go to the production server. If server~3 is unavailable, queries go to server~2 instead. Load-balancing or additional servers are not necessary at this point; however, this could be trivially implemented if and when it becomes necessary.  Server~1 does not accept regular user queries but could be opened in case the two other servers become unavailable.

Server~3 also runs the PostgreSQL database. The majority of analysis tasks do not require access to the slow control data so that temporary loss of access has minor impact on the project and it has not been necessary to provide a fall-back server. Since the DeltaV system itself acts as a backup to the slow control data, mirroring the data on another server as backup is not necessary.

\section{The CouchDB database}\label{sect:CouchDB}
Couch\-DB is a system for storing JSON documents. It is well suited for situations where many disparate types of data must be stored, and where the structure of the data is expected to change with time. For example, when a new quantity is introduced to calibration, this quantity can be stored from a certain time on without impacting the existing documents.

\subsection{Requirements}
The PMT data is reduced automatically on high performance computing clusters in close to real time. With the detector operating at over 95\% uptime, this leads to a baseline Couch\-DB load of approximately 84 read and 24 write operations per hour. In addition to this baseline, the database must support a peak load of $\mathcal{O}(10^3)$ simultaneous read requests. This typically happens when a large number of runs is re-analysed simultaneously with newer software, and during Monte Carlo simulation campaigns. 

Occasional bursts of write-operations on top of the baseline occur approximately once per month, when calibration constants for many runs are uploaded. These are of $\mathcal{O}(100)$ per second. In nearly all situations, data is uploaded at least several minutes before it is first read back. Overall, access to this DB is heavily biased toward read operations.

We project the total number of database documents by the end of the project to be $\mathcal{O}(10^5)$.

These requirements are well within the design specification for the Couch\-DB database system.

\subsection{Data structure} \label{sec:datastructure}

When we refer to \emph{data} from here on, we mean non-event information stored in one of the databases. We will use \emph{PMT data} to denote the waveform data from the PMTs.

Data is stored in Couch\-DB as JSON objects. Each JSON object is called a `database document' or dbdoc\footnote{While design documents are also JSON objects, we specifically mean regular documents here.}. Couch\-DB imposes no further rules on the data structure.

RAT imposes a set of common keys on JSON files read from the local disk, which structure the data into namespaces and run ranges. To maintain compatibility, we impose this structure on the remote dbdocs. RAT-DEAP introduces additional keys which support accountability and revision control. The required and optional keys are listed in Tab.~\ref{tab:commonkeys}. The actual data is then stored under a number of additional keys. We denote those keys \emph{fields}. 
A valid document would be for example\footnote{We are not including the Couch\-DB-internal keys \_id and \_rev .}:

\begin{minted}[fontsize=\footnotesize]{json}
{ "name":"PMTGAIN", 
  "index":"100", 
  "run_range":[17600, 17800], 
  "author":"Tina Pollmann", 
  "createdOn":1469112443, 
  "RATVersion":"v5.1.7-242-g35f643d", 
  "SPE":10.0 }
\end{minted}

\begin{table}[htb]
\caption{Mandatory (above the horizontal line) and optional (below the line) keys. }
\begin{center}
\rowcolors{3}{tablerowcolor}{}
\begin{tabular}{L{0.01\linewidth}|L{0.2\linewidth}|L{0.62\linewidth}}
\rowcolor{tableheadcolor}
  & Key name & Purpose \\
\hline
\cellcolor{white!0}  &name & Data is separated into namespaces. This is the top level namespace. \\
\cellcolor{white!0}  &index & Second level namespace, differentiates between tables of the same \textbf{name}. This is typically an object identifier from Tab.~\ref{tab:tagnames}, such as the PMTID.  \\
\cellcolor{white!0}  &run\_range & The first and last RunID for which this document is valid. \\
 \cellcolor{white!0} &author & The name of the person who wrote this data to the DB.  \\
\cellcolor{white!0}\multirow{-5}{*}{\rotatebox[origin=c]{90}{mandatory}}  &createdOn & POSIX time stamp when this document was added to the database. \\
\hline
 \cellcolor{white!0}   & notes & Any other relevant text information about the data in this document. \\
 \cellcolor{white!0}   & RATversion & For calibration results, the Git tag of the software version used to obtain the result. \\
\cellcolor{white!0}    & old & A structure that keeps older revisions of the data in this document. \\
\cellcolor{white!0}\multirow{-3}{*}{\rotatebox[origin=c]{90}{optional}} &deprecatedOn & POSIX time stamp when this data was deprecated.
\end{tabular}
\end{center}
\label{tab:commonkeys}
\end{table}%

Typically, documents with the same \textbf{name} have the same fields, but this is not a strict requirement.

CouchDB allows dbdocs to have binary attachments. For example, ROOT files or png image files can be attached to a dbdoc. Such dbdocs have a key with a string-type value that contains the name of the attached file. This name is used by RAT-DEAP to retrieve the file and make it available to the user.

As foreseen by the RAT architecture, the basic unit of time in this database is the run. All information is valid only for a certain range of RunIDs. The group of dbdocs with the same \textbf{name} and \textbf{index} but different run ranges is referred to as a \emph{data-group}. Within the same data-group, run ranges must not overlap.

Normal \textbf{run\_range} keys are in the form of [n, m], where n and m are RunIDs. The RunIDs \SI{-2}{}, \SI{-1}{} and 0 have special meaning. A RunID of \SI{-2}{}, which is specific to RAT-DEAP, indicates an open ended validity range. In other words, a \textbf{run\_range} of [n, \SI{-2}{}] makes the dbdoc valid for RunID $\geqslant$ n. A RunID of \SI{-1}{} is used for user overwrite. If any of the dbdocs within a data-group have a run validity of \SI{-1}{}, this dbdoc is always returned. This feature is not used in the central production database but is useful for testing. A RunID of 0 indicates a default dbdoc. If any dbdoc in a data-group has RunID 0, this dbdoc is used if data valid for the target run is not available.

The presence of the mandatory keys is enforced through Couch\-DB's \emph{validate\_doc\_update} function. The mandatory keys can be considered metadata to the data that is actually of interest in the dbdocs. No rules are imposed on how the rest of the dbdoc is structured.

Many of the constants saved in the database make sense only if they exist for all items in a set, such as all PMTs or all DAQ channels. In the default RAT, parameters for such sets of objects are always saved as an array, where the array index corresponds to the object ID. A document could for example look like (array truncated after 5 objects):
\begin{minted}[fontsize=\footnotesize]{json}
{ "name":"PMT",  "index":"gain", 
  "run_range":[17600, 17800], 
  "SPE":[10.0, 10.3, 9.5, 9.9, 10.4] }
\end{minted}

However, particularly for PMT calibration constants, we decided to create separate dbdocs for each PMT. The document \textbf{index} is the ID of the PMT the document pertains to.  For example, documents with \textbf{name} ``PMTSPE'' contain PMT gain parameters, and the data-group of \textbf{name} ``PMTSPE'' and \textbf{index} ``123'' contains the gain parameters for PMTID~123. This data-group is subdivided into run ranges as needed. The advantage of this scheme is that a PMT with stable gain may be described by one dbdoc with a large run range, while a PMT with drifting gain can have several dbdocs with shorter run ranges.

This scheme makes sense when parameters are valid over many runs, and the validity range is different for different items in the set. This scheme makes no sense for parameters that are determined on a run-to-run basis, and that need to be stored for each single run, such as the bias voltage on each PMT. In this situation, an array field is added to the run configuration dbdoc, as forseen in the default RAT.

\subsection{Finding the right data} \label{sect:findingdata}
The Couch\-DB map/reduce system is used to sort and select data from the databases. Each database within the Couch\-DB system has two design documents. Within Couch\-DB, design documents are dbdocs whose name starts with `\_design'. They contain instruction for sorting the data in the regular dbdocs. These instructions are called \emph{views}. One design document is general purpose and the other specifically defines the interface to RAT-DEAP. The views in the general purpose design document support the webpage display of database entries, as well as queries from Python programs. Because they are in their own design document, they can be modified and extended without affecting the RAT interface.

To maintain compatibility with using JSON-like text documents from the local system, RAT-DEAP always fetches and reads in full JSON objects. Data retrieval does not rely on views specific to certain types of data, but uses just one view to identify and download the document identified by a particular name and index which is valid on a target run. All fields in that document are then made available to the user. This dbdoc selection is achieved by a single view:

\begin{minted}[fontsize=\footnotesize]{javascript}
"select": { 
  function(doc) { 
     if (doc.hasOwnProperty("name") 
       && doc.hasOwnProperty("index") 
       && doc.hasOwnProperty("run_range")) { 
         emit(
              [doc.name, doc.index, doc.run_range[0]], 
              doc.run_range[1] 
              );
    }
  } 
}
 \end{minted}

This \emph{select} view sorts all the dbdocs first by their name, then by their index, and last by their start RunID n.

In analysis code, the user specifies the name and index of the data-group that contains the field to read data from. RAT-DEAP constructs a Couch\-DB query to the \emph{select} view:
\begin{minted}[fontsize=\footnotesize]{javascript}
    _view/select?startkey=[name,index,RunID]
                &endkey=[name,index,endrun]
                &descending=true
                &limit=1
                &include_docs=true
\end{minted}   
where the options after the question mark are query parameters. The RunID of the target run is supplied by RAT-DEAP at runtime. The `endrun' variable used in the above code snippet is set to mind the special validity ranges \SI{-1}{} and 0:
\begin{equation}
\text{endrun} = \begin{cases}
\text{RunID} &; \text{RunID} \leq 0 \\
1 &; \text{RunID} > 0 
\end{cases}
\end{equation}  
   
The document returned by the \emph{select} view is guaranteed to be from the correct data-group, and within the data-group has the biggest n for which n$\leq$RunID. The requirement that (RunID$\leq$m if m$>$0) is checked by RAT-DEAP. RAT-DEAP also handles the logic that deals with user-overwrite and default dbdocs.

To illustrate the retrieval mechanism, consider the following set of sample dbdocs:
\begin{minted}[fontsize=\footnotesize]{json}
{ "name":"PMTGAIN", "index":"100", 
              "run_range":[0, 0], "SPE":10.0}
{ "name":"PMTGAIN", "index":"100", 
              "run_range":[10, 19], "SPE":11.5}
{ "name":"PMTGAIN", "index":"100", 
              "run_range":[20, 34], "SPE":11.7}
{ "name":"PMTGAIN", "index":"100", 
              "run_range":[35, 75], "SPE":11.9}
\end{minted} 

The \emph{select} view for these documents without query parameters returns (irrelevant fields omitted):
\begin{minted}[fontsize=\footnotesize]{json}
[{ "key":["PMTGAIN","100", 0], "value":0  },
{ "key":["PMTGAIN","100", 10], "value":19  },
{ "key":["PMTGAIN","100", 20], "value":34  },
{ "key":["PMTGAIN","100", 35], "value":75  }]
\end{minted} 

A user is analyzing RunID 23 and requests a field from this data-group (name ``PMTGAIN'' and index ``100''). With the \emph{startkey}, \emph{endkey}, and \emph{descending} query parameters the view result is:
\begin{minted}[fontsize=\footnotesize]{json}
[{ "key":["PMTGAIN","100", 20], "value":34  },
{ "key":["PMTGAIN","100", 10], "value":19  },
{ "key":["PMTGAIN","100", 0], "value":0  }]
\end{minted} 

The \emph{limit} query parameter then selects the first row of the result: 
\begin{minted}[fontsize=\footnotesize]{json}
[{ "key":["PMTGAIN","100", 20], "value":34  }]
\end{minted} 

The \emph{include\_docs} parameter causes the full document that belongs to this row to be returned with the view result, so that no further database queries are necessary.

A typical analysis often needs the same type of information for a whole set of objects, such as the gain parameter for each PMT. This information is located in dbdocs with different indexes but the same name. The above scheme can be used to make many subsequent network queries to obtain each dbdoc one after the other. This can be slow, especially if the network connection is unreliable. The same result can be achieved with only two network requests. First, all rows that belong to dbdocs with the specified name, regardless of run range or index, are selected by querying
\begin{minted}[fontsize=\footnotesize]{javascript}
  _view/select?startkey=[name,0,0]
                &endkey=[name,{}]
                &include_docs=false
\end{minted}   
The dbdoc contents are not requested at this stage (\emph{include\_docs=false}). RAT-DEAP now loops over the rows in the view result and saves the IDs of dbdocs with the correct run validity in an array we call \emph{selectedids}. All those dbdocs are then fetched at once by querying:
\begin{minted}[fontsize=\footnotesize]{javascript}
   _all_docs?keys=selectedids&include_docs=true
\end{minted}
This requires more logic to be implemented in RAT-DEAP, and the initial view result can be fairly large, but in most situations, the gain in speed and reliability is worth the effort.

\subsection{Revision control}
Sometimes, calibration constants or other metadata change after they were added to the database and used in analysis. The goal of revision control (RC) is to make it possible to retrieve the value that was valid for a specific run at an earlier date. Since that earlier date, the value, the run range for this value, or both, could have changed. 
Couch\-DB has no built-in RC features, so this must be implemented on the user side.

\begin{figure*}[htbp]
\begin{center}
\includegraphics[width=0.7\textwidth]{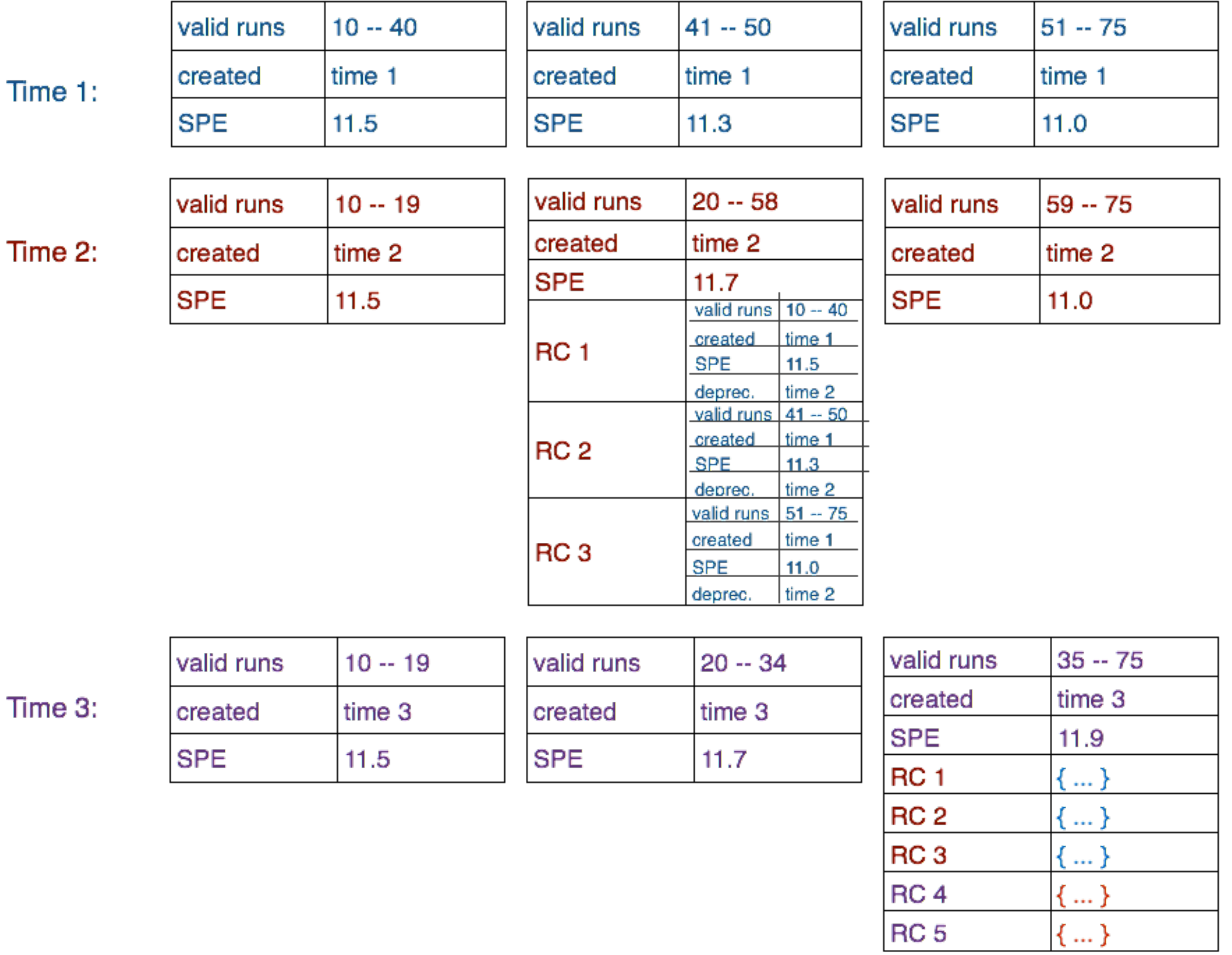}
\caption{Illustrative example of the revision control scheme implemented for Couch\-DB documents. Each table represents one dbdoc. The three rows show the database state for a data-group at different times. dbdocs are originally entered at time 1 (top row). The field of interest and the run validity are changed once at time 2 and again at time 3. Each time, the original dbdocs are saved in their entirety as a nested structure within one of the new documents, and the `deprecated on' (abbreviated as `deprec.' here) key is added. Information added at time 1, 2, and 3 is colored blue, red, and purple, respectively. In the `time~3' row, the information kept in the RC structures is not written out. The RC~1 through 3 fields are the same ones as in the middle dbdoc in the `time~2' row. The RC~4 and 5 fields are created following the same rules.}
\label{fig:dbrevcontrol}
\end{center}
\end{figure*}

For concreteness, consider a situation where the gain (in the form of a single photoelectron, or SPE, charge) for PMT 30 was determined at time T1 to be 11.5~pC (run 10 through 40), 11.3~pC (run 41 through 50), and 11.0~pC (run 51 through 75). This database state is shown in the top row of Fig.~\ref{fig:dbrevcontrol}.

At a later time T2, a new analysis determined that the SPE charge was really 11.7~pC from runs 20 through 58. The new state is shown in the second row of Fig.~\ref{fig:dbrevcontrol}. To make this change, the run range of two dbdocs had to be modified. One dbdoc was deleted and replaced by a new one containing the new field value and new run range. New RC fields of JSON-type value are added to this dbdoc. These fields each contain one of the three original dbdocs.

The map/reduce function shown earlier is blind to the RC fields, so that in regular use, only the current state of the database is exposed to the user. However, it is possible to return data from a specific date by querying a second view, which loops over the RC structures:
\begin{minted}[fontsize=\footnotesize]{javascript}
"selectold": { 
  function(doc) { 
   if (doc.hasOwnProperty('name') 
         && doc.hasOwnProperty('index') 
         && doc.hasOwnProperty('run_range')) { 
     emit([doc.name, doc.index, doc.run_range[0]], 
           doc.run_range[1]);
     int nrev = 1;
     while( true ) {
      if (doc.hasOwnProperty('old_' + nrev) {
         emit([doc.name, doc.index, 
               doc['old_' + nrev].run_range[0]], 
               doc['old_' + nrev].run_range[1]);
      }
      else break;
      nrev = nrev+1;
    }
  }
}
 \end{minted}   
This view returns the database state at all previous times. RAT-DEAP then checks which of those documents was valid on the date given. 

Many rows in the view result could now represent the document that contains the desired value so that the query parameters can no longer limit the answer to just one document. In general, we cannot know how many documents might be relevant so we should not limit the result. However, in the scope of this project, values are not updated very frequently. No more than $\mathcal O (10)$ entries are relevant, so that the number of returned documents can be limited to 20--50.

\subsection{Security} \label{sec:security}

The goal of the security system is to manage who can edit what data in the database, so that the risk of both accidental and malicious changes is minimized.

We define several Couch\-DB user accounts with permissions that are limited through Couch\-DB's \emph{validate\_doc\_update} function to specific fields and data-groups. A Couch\-DB `user' account can be shared by several people. Accounts exist for the deapdb managers, DAQ experts, the DAQ operators, the analysis coordinator, the run coordinator, and the data quality coordinator.

The deapdb manager has superuser access with add/\-edit/\-delete permissions to all dbdocs in deapdb. The DAQ experts account has management access to daqdb and write access to specific fields in deapdb. Likewise, the run coordinator has management access to scheduledb and write access to specific fields in deapdb.

The document update function forbids add/\-edit/\-delete operations of regular dbdocs by the server administrator, and forbids modification of two types of documents even by the database managers\footnote{Though the db managers and server administrator could modify the document update function to give themselves edit permission to these documents.}: dbdocs that describe a DAQ shift cannot be modified after the date of the shift. Runlists cannot be modified after analysis results based on them have been published.

\subsection{Offline operation}
Couch\-DB packages are available for all major operating systems. Users can replicate the official database in whole or in parts to a Couch\-DB instance installed locally. RAT-DEAP is set up to automatically connect to a local Couch\-DB server if the remote servers cannot be reached.

Alternatively, all or select dbdocs can be downloaded as individual JSON files into a local directory. Pointing RAT-DEAP to this directory, it will read in these documents and retrieve their data. This slows down RAT-DEAP startup considerably, and does not support binary attachments to dbdocs, which are needed for some types of analysis, but is a viable option for systems where neither installing user software nor remote queries are possible.

\section{The PostgreSQL database} \label{sect:dpostgres}
Analysis-relevant slow control sensor data accumulated by the DeltaV system is continuously exported to the PostgreSQL server so that users can access this data while insulating the DeltaV system, which controls the detector's gas handling and cooling systems. The PostgreSQL server is queried using a custom-written HTTP interface. The query contains the sensor name and a time range, and the interface responds with a text object containing arrays of time-stamps and the corresponding sensor readings.

The time-ordered readings from the 127 analysis-relevant sensors are stored in a fixed scheme of [sensor-tag, timestamp, float value]. Data from each sensor is written to the DB every \SI{30}{\second} and the DB is read approximately 5 times a day during standard detector operation and analysis. Because the data scheme is not expected to change, and this database is heavily biased toward write operations, PostgreSQL is a system more suitable than Couch\-DB for this subset of the metadata.

Most user queries for non-event data are mediated through dedicated classes in RAT, and the two database-system backend solution is not exposed to most regular users. For example, a user wanting to know the detector pressure during a given run uses the same class they also use to find out the PMT voltages, or to check if a calibration source was deployed.

Since only one system ever writes to this database, only one user with edit rights exists. No revision control is implemented, as we do not expect to update sensor readings after they are recorded.

\section{Interfaces}

Data flows in and out of the databases through the RAT-DEAP (C++/Python) analysis framework, through the database web interfaces, and through Python scripts. 

The vast majority of database queries are performed through RAT-DEAP. Most of the objects from Tab.~\ref{tab:tagnames} exist in RAT-DEAP as classes which hide the underlying database implementation from the user, so that the way the databases store information can be changed without requiring updates to user code. For example, in order to find out which PMTID a certain optical calibration source is installed on, the user would use the function `PMTIDforSourceID(int sourceID)' provided by RAT-DEAP, rather than construct their own database query. RAT-DEAP directs queries either to Couch\-DB or to PostgreSQL as needed.

\begin{figure}[htbp]
\begin{center}
\includegraphics[width=0.48\textwidth]{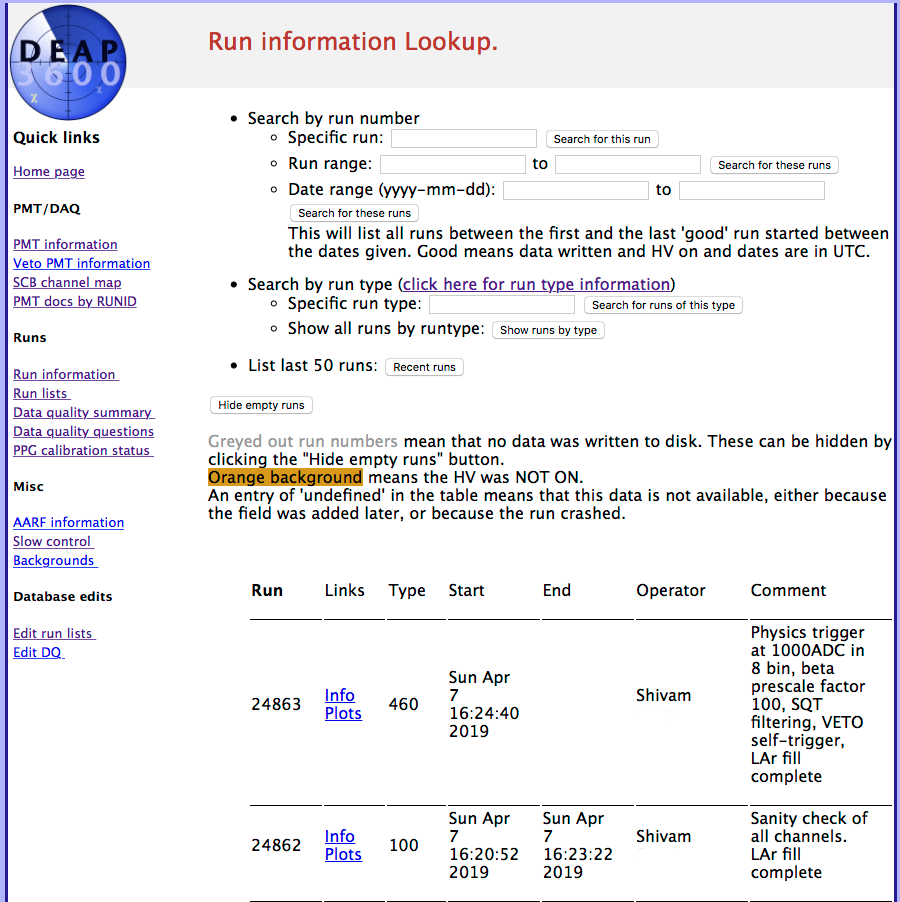}
\caption{Example webpage from the DEAP database WebView. The webpage displays information about runs.}
\label{fig:webview}
\end{center}
\end{figure}

We built a CouchApp on Couch\-DB and a website on top of PostgreSQL. Most information in the databases is surfaced in a user-friendly way on these websites. An example site is shown in Fig.~\ref{fig:webview}. Certain fields in Couch\-DB can also be edited from the website.

Python libraries exist for performing Couch\-DB and HTTP queries. Python scripts are used extensively to manage the databases, and for all tasks related to managing the analysis-processes of the datasets.

\section{Workflows} \label{sect:workflow}
\subsection{DAQ integration}
\begin{figure}[htbp]
\begin{center}
\includegraphics[width=0.48\textwidth]{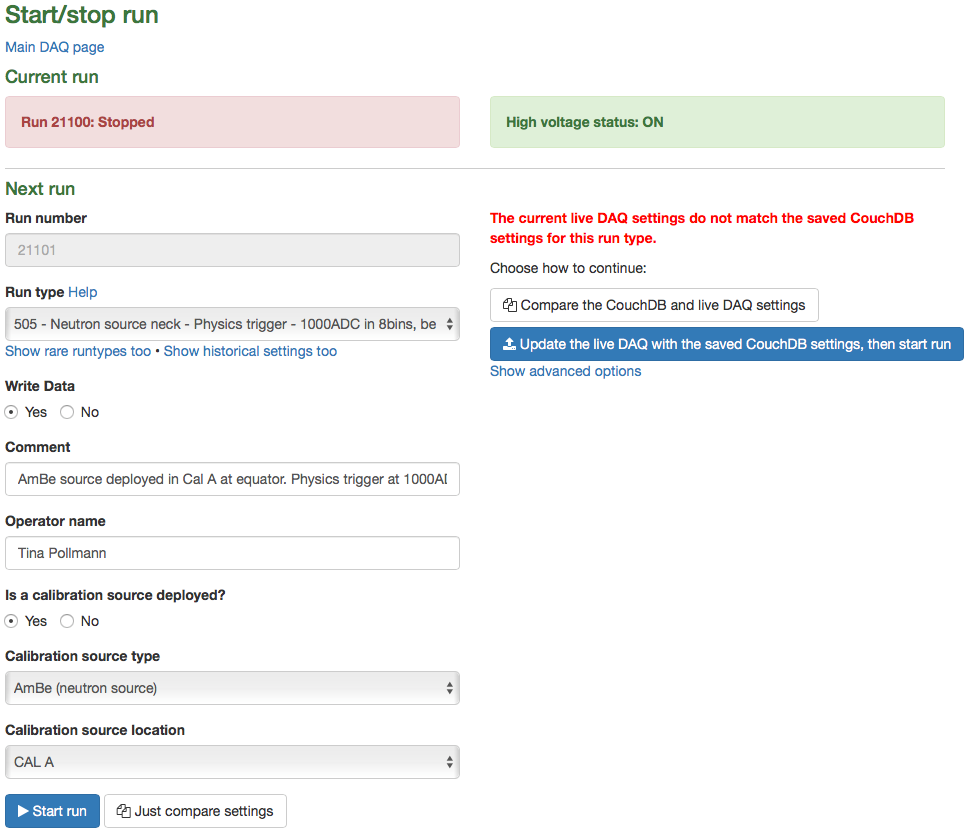}
\caption{The MIDAS web interface for starting a new run.}
\label{fig:newruninterface}
\end{center}
\end{figure}

DEAP uses the MIDAS~\cite{midas} system to manage data acquisition. The analysis goal for a dataset determines how MIDAS is configured. For example, a run to monitor dark noise is set up differently from an optical calibration run, and both are different from a run collecting physics data. MIDAS was extended for DEAP to support importing and exporting the MIDAS configuration in JSON format. In order to ensure that runs of a specific type are always taken in exactly the same DAQ configuration, these configuration files defining the run types have unique IDs and are managed by Couch\-DB. When starting a new run, the operator selects the ID of the desired run type. The DAQ queries Couch\-DB for the configuration file, applies the settings stored in the file, then starts the run. The interface is shown in Fig.~\ref{fig:newruninterface}.

In addition to the run type, the operator provides his or her name, a run comment, and information on calibration sources in use when applicable. This information is assembled into a new JSON document and uploaded to deapdb. At the end of the run, the DAQ updates this document with information such as the length of the run and the number of data files created. A full export of the DAQ settings at the start and at the end of the run is also attached to the run document\footnote{MIDAS allows some settings to be changed while a run is ongoing, hence the settings at start and end of run are compared.}.

At minimum, four runs are taken per day; three to verify calibrations, and one physics-data run. Every change in configuration results in a new run. For example, a full optical calibration campaign goes through approximately 10 optical calibration sources at 10 intensities each, resulting in \SI{100}{} runs taken within one day.

A number of automated or semi-automated scripts periodically query \emph{deapdb} for new runs. They initiate transfer of PMT-data files to analysis clusters and permanent storage sites, and launch automated data reduction and calibration routines.

\subsection{Data quality and run selection system}
A data quality (DQ) document is automatically created for each run. Data quality is evaluated at 4 levels:
\begin{itemize}
	\item Level 1: Automated checks done by MIDAS at end of run.
	\item Level 2: Standard DQ checks by DAQ operator at end of run.
	\item Level 3: Standard DQ checks based on offline analysis results.
	\item Level 4: Assessment by data quality working group.
\end{itemize}

The checks for level~2 and level~3 are performed by answering questions about online (level~2) and offline (level~3) plots. Online plots are populated by the DAQ during the run and provided to the shifter for assessment at the end of the run. Offline plots are created during the initial automatic processing by RAT-DEAP and include values calculated based on the PMT data as well as slow control data read from the PostgreSQL database. 

The DQ doc stores the answers to the DQ questions, and a DQ summary flag for each level. It also contains the run narrative: This is a time stamped set of text strings written by the DAQ operator to describe things that happened during the run and that might be of interest to analyzers.

With approximately \SI{20000}{} runs expected by the end of the data-taking phase (over \SI{10000}{} calibration and testing runs were recorded before the detector was even filled with the liquid argon target), manual selection of datasets is unrealistic. Datasets are created semi-automatically by a Python program, based on input criteria such as the run date, run type, run duration, and data quality flags. The datasets thus assembled are saved as \emph{runlists}. In addition to the standard fields and the array of runs, the dbdocs save the run-request ID\footnote{Collaboration members requesting data to be taken in a specific new configuration issue a \emph{run request}. To find their data, they can then search the database for runlists related to their run-request ID.} and a comment about the intended analysis use of this list. Runlists are locked when a result is published, to preserve the information about which exact set of runs were used for that analysis.

\subsection{DAQ shifts}

DAQ operators do remote shifts in a 24-hour rotation, and there are requirements on the number of shifts covered by each member institution per year.

A CouchApp facilitates scheduling of shifts and reporting of shift statistics. The name, contact information, institution, and status (whether they are active or retired) of each DAQ operator is stored in the DB. DAQ operators indicate days when they are available for a shift using the \emph{scheduledb} web interface. This creates shift documents which contain the shifter information, the shift date, the shift credit, and the scheduling status. 

Once per week, the run coordinator checks the calendar on the same web interface and assigns the shifts. 

Statistics about assigned shifts, such as the fraction of shifts covered by each institution in a given month or year, are displayed live in the CouchApp.


\subsection{Detector response calibration}
Deapdb tracks detector response parameters at run or subrun-level\footnote{The data of a run is distributed over many files, such that each file has a fixed size, as some computing systems cannot handle very large files. This divides each run into subruns.}  granularity. Some parameters, such as the time synchronisation of DAQ channels, are determined in initial RAT-DEAP processing of a subrun and automatically uploaded to Couch\-DB. Other parameters, such as PMT gains or dark noise levels, are extracted periodically from calibration runs and the resulting JSON files are manually uploaded to Couch\-DB.

\section{Performance} \label{sect:performance}


Four years after the PMTs were first turned on, the deapdb database has \SI{164983}{} documents. Of these, 67\% pertain to PMT response, 25\% pertain to runs, and 7\% pertain to DAQ settings and response. The remaining 1\% of documents contain information on other miscellaneous things, such as slow control sensors and calibration sources.

The main user server receives on average 30 HTTP GET requests per second. These lead to on average 475 dbdocs read per second. This request load varies strongly with time. During large simulation or data reduction campaigns, up to \SI{1648}{} GET requests per second have been reached.

Request processing times are strongly dependent on the system the database runs on. However, we can compare the RAT-DEAP implementation of the \emph{select} view to that in the default RAT: the implementation in RAT-DEAP makes the Couch\-DB view index approximately 5 times faster to build from scratch and takes $\mathcal O (10^4)$ times less storage space on the server. On the DEAP default server, a request takes on average 48~ms to process. However, this can increase to several seconds if Couch\-DB has to re-build a view. 

Some data-groups have been surpassing \SI{50000}{} documents. Requests selecting documents within these groups can take up to several minutes at times when Couch\-DB is under peak load (see Sect.~\ref{sect:findingdata}). In the future, optimization in how documents are retrieved in this situation will be necessary, for example by creating custom Couch\-DB views for some types of metadata.

Fewer than 0.1\% of analysis jobs submitted to the computing clusters fail due to connectivity problems to the database. Such failure is most often due to network problems of nodes on the computing clusters. 

\section{Discussion}

By using Couch\-DB for most types of non-event data, we have emphasised flexibility over speed and efficiency. During the first two years of operation, the non-event data stored, the document structure, and the view structure was changed frequently without impacting the ongoing analysis efforts. This allowed us to optimise the database usage and adapt it to emerging requirements.

Couch\-DB guarantees so-called \emph{eventual consistency} for replications such as those discussed in Section~\ref{sect:infrastructure}. This can lead to race conditions, where, for example an automatic process on the DAQ writes into a data quality dbdoc on Server~1 while at the same time a user writes into the same dbdoc using the web interface on Server~2. Couch\-DB will pick a winning document automatically but note that there is a conflict. Depending on which edit wins, either the information from the DAQ system or the information from the user is not available until the conflict is resolved. Most of the time, resolving the conflict is trivial and can be done by a dbdoc merge script. However, the database managers have to make sure conflicts do get resolved. 

Eventual consistency could become an issue in experiments that routinely have to read back information that was written a short time earlier to a different mirror. Under normal conditions, information is replicated within minutes between the three servers.

The decision to store the calibration constants for each PMT in an individual dbdoc (see Section~\ref{sec:datastructure}) was based on the same consideration of flexibility. For experiments using more than $\mathcal O (100)$ light detectors or similar hardware units, tracking of their properties using the document structure described here will no longer be the best option. In that case, storing constants for all the hardware units in an array within a single document is likely the better design choice.

We optimised the view by which RAT-DEAP finds documents in the database, but kept the general architecture where RAT always deals with full JSON documents. This means that documents managed by Couch\-DB and those that exist locally within the RAT install can be processed to extract their data and cache the information in the same way. By doing this, we treat Couch\-DB as nothing more than a store of JSON objects. The speed of DB requests could be improved significantly if additional named views, specific to certain often-used calibration constants, were implemented. RAT-DEAP would then fetch only the constant or array of constants needed, not the complete JSON document that contains the constant(s). 

This mode of retrieving data is already used for lookups in the PostgreSQL database, which does not contain JSON documents. The whole JSON document is never directly exposed to the user --- users always work with single constants or arrays of constants which RAT-DEAP either extracts from a JSON document or from PostgreSQL entries --- so switching to a named-view based retrieval mechanism will not break user code and can be implemented at any time needed without disturbing ongoing analysis.

\section{Conclusion}
We have built a non-event data store for the DEAP-3600 dark matter detector to support data analysis and detector operation.

For an analysis database to be useful, the data in it must be complete and correct. To meet the physics goals of the project, this non-event data must be readable by the data analysis software at runtime. It must be available from a central location so that all data analysts use the same, most up-to-date, non-event data. It must be easy to access and be accessible at all times, so that analysts can work efficiently. It must be available through different interfaces because the diversity in analysts and analyses that are part of a  project like DEAP makes it unrealistic and undesirable to lock anyone into a specific access scheme. The source of the metadata must be obvious so that if questions about its validity arise, the workflow that resulted in the data entry in question can be reproduced and verified. Some metadata must be revision controlled, with the option to access older versions of an entry if desired, so that the results of data analysis relying on this data remain reproducible.

We have achieved all these requirements using Couch\-DB as the non-event data store, with PostgreSQL for specific highly-ordered types of the non-event data. We have built on the database scheme that is part of the standard RAT install, and implemented both a Couch\-DB and PostgreSQL backend. User requests for data are routed to the correct database automatically.

Flexibility and speed of deployment were crucial at the beginning of the project. As the non-event data grows and the project matures, the efficiency of lookups can be improved without disturbing ongoing analysis efforts or breaking older user code. 

On the detector operation side, a project acquiring data continuously for many years needs automated processes to sort, store, and analyse the datasets. Information describing datasets is the second most common type of information we store in Couch\-DB. It enables these automated processes to operate with little user input. It also supports transparency by allowing all collaboration members to access real-time or near real-time statistical information on all aspects of data taking.

\section{Acknowledgements}

We would like to thank the DEAP collaboration for suggestions and bug testing. Particular thanks go to Mike Hamstra for making the dataflow from the DeltaV system to PostgreSQL system work, and for his operational support of the database servers. We also want to thank the original authors of RAT, in particular Stan Seibert, for devising an excellent software architecture that allowed for straight-forward additions of the remote database backends.

\bibliography{dbbibcomplete}
\bibliographystyle{tp_unsrt_doi}
\end{document}